\documentclass{emulateapj}
\usepackage{graphicx}
\usepackage{epstopdf}
\begin{document}

\title{Ocean Planet or Thick Atmosphere: On the Mass-Radius Relationship for Solid Exoplanets with Massive Atmospheres}

\author{E. R. Adams\altaffilmark{1}, S. Seager\altaffilmark{1,2}, L. Elkins-Tanton\altaffilmark{1}}

\altaffiltext{1}{Department of Earth, Atmospheric, and Planetary Sciences, Massachusetts Institute of Technology, 77 Massachusetts Ave., Cambridge, MA, 02139}

\altaffiltext{2}{Department of Physics, Massachusetts Institute of Technology, 77 Massachusetts Ave., Cambridge, MA, 02139}

\begin{abstract}

The bulk composition of an exoplanet is commonly inferred from its
average density. For small planets, however, the average density is
not unique within the range of compositions. Variations of a number of
important planetary parameters---which are difficult or impossible to
constrain from measurements alone---produce planets with the same
average densities but widely varying bulk compositions. We find that
adding a gas envelope equivalent to 0.1\%-10\% of the mass of a solid
planet causes the radius to increase 5-60\% above its gas-free
value. A planet with a given mass and radius might have substantial
water ice content (a so-called ocean planet) or alternatively a large
rocky-iron core and some H and/or He.  For example, a wide variety of
compositions can explain the observed radius of GJ 436b, although all
models require some H/He. We conclude that the identification of water
worlds based on the mass-radius relationship alone is impossible
unless a significant gas layer can be ruled out by other means.

\end{abstract}


\section{Introduction}

Out of over 250 exoplanets known to date, over 20 are known to transit their
stars. Transiting planets are important because we can derive the
precise mass and radius, and can begin to determine other planetary
properties, such as the bulk composition.

Much attention has been given to ``ocean planets'' or ``water
worlds'', planets composed mostly of solid water \citep{kuch2003,
lege2004}. If a water world is found close to a star, it would be
strong evidence for migration because insufficient volatiles exist
near the star for {\it in situ} formation. The proposed identification
of water worlds is through transits. From a measured mass and radius a
low-density water planet could potentially be identified.

We examine the possibility that water worlds cannot be uniquely
identified based on the mass and radius of a transiting planet.  An
alternative interpretation could be a rocky planet with a thick
hydrogen-rich atmosphere.  Most authors have assumed that solid
planets in the 5 to 10 $M_{\oplus}$ range have an insignificant
amount of hydrogen \citep{vale2006, vale2007a, fort2007, seag2007,
sels2007, soti2007}.

Exoplanets have, however, contradicted our basic assumptions before.
Notable examples include: the existence of hot Jupiters; the
predominance of giant planets in eccentric orbits; and the gas-rock
hybrid planet HD~149026b with its $\sim 60 M_{\oplus}$ core and $\sim
30 M_{\oplus}$ H/He envelope \citep{sato2005}.

We adopt the idea that a wide range of atmospheric formation and loss
mechanisms exist and can lead to a range of atmosphere masses on
different exoplanets. We explore the mass-radius relationship for the
lowest-mass exoplanets yet detected ($\sim$~5-20 $M_{\oplus}$) in
order to identify potential ambiguities that result from the presence
of a massive atmosphere. We explore atmospheres ranging from $\sim
10^{-3} M_{\oplus}$ (10 $\times$ Venus' atmospheric mass) to $\sim 1
M_{\oplus}$ (the estimated mass of Uranus' and Neptune's H/He \citep{guil2005, podo1995, hubb1991}), with a
focus on the smaller mass range. We also explore potential
compositions for the transiting Neptune-size planet GJ~436b
\citep{butl2004, gill2007a}.

\section{Models}

Our model assumes a spherical planet in hydrostatic equilibrium, with
concentric shells of different composition sorted by density. We solve
for the mass $m(r)$, pressure $P(r)$, density $\rho(r)$, and
temperature $T(r)$ using the equation for mass of a spherical shell, 
\begin{equation}
\frac{dm(r)}{dr} = 4 \pi r^2 \rho(r),
\end{equation}
the
equation of hydrostatic equilibrium, 
\begin{equation}
\frac{dP(r)}{dr} = -\frac{G m(r) \rho(r)}{r^2},
\end{equation}
and the equation of state of the
material. 
\begin{equation}
\rho(r) = f(T(r),P(r)).
\end{equation}

To build a model planet, the composition and mass of up to three
layers are specified. Starting from a central pressure adjusted to
achieve the desired mass, the differential equations for $P(r)$ and
$m(r)$ are numerically integrated outward, and $\rho (r)$ is
calculated from the equation of state for the material. At the
boundary between layers, $P(r)$ and $m(r)$ are used as the initial
conditions for the next layer.  The integration is stopped at the
outer boundary condition $P(r) = 1$~bar, approximately where the
atmosphere ceases to be opaque at visible and infrared
wavelengths\footnote{We ignore any wavelength-dependent effects on
potential measured radii of extrasolar planets.}.

The temperature of a planet's gas layer has a significant effect on
the radius. Instead of computing a cooling history to obtain $T(r)$,
we calculate $T(r)$, separately in three different regimes. Within the
solid portion of the planet ($r \geq r_{solid}$), temperature has
little effect on the final radius, and is assumed to be isothermal
\citep[see][]{seag2007}.  In the deep hydrogen-helium layer where the
pressure is greater than 1 kbar ($r_{solid} \leq r \leq
r_{1\vspace{0.1in } {\rm kbar}}$), we assume that convection dominates
and the temperature follows an adiabat tied to the entropy, $S$, at 1
kbar. At lower pressures ($r > r_{1\vspace{0.1in}{\rm kbar}}$), we use
the radiative equilibrium gray analytical model of \citet{chev2007}
and \citet{hans2007} for irradiated atmospheres,
\begin{equation}
T(r) = \frac{3}{4} T_{eff}^4 \left(\tau + \frac{2}{3} \right) +
T_{eq}^4 F(\tau, \mu_0, \gamma),
\end{equation}
where $\tau$ is the optical depth, $\mu_0 = \cos \theta_0$ where
$\theta_0$ is the angle of incident radiation with respect to the
surface normal, and $\gamma$ is a parameter that accounts for the
altitude at which radiation is absorbed.  We convert from optical
depth to pressure through a constant scaling relation under the
assumption of hydrostatic equilibrium and constant opacity per gram,
following \citet{hans2007} but taking into account the planet's
surface gravity.  Here $T_{eff}$ is the effective temperature in the
absence of stellar irradiation, representing energy from internal
sources. $T_{eq}$ is the equilibrium temperature which represents
heating from the parent star. 

We emphasize that by choosing an effective temperature---or
adiabat---we are subsuming a cooling calculation. This is common
practice for modeling the interior structure of solar system giant
planets \citep[e.g.,][]{stev1982, hubb1991, marl1995, Saum2004} and
was also the case even before their $T_{eff}$s and gravitational
moments were known or well known \citep{dema1958, zs1969}.  In fact,
even with a known $T_{eff}$ cooling calculations only match observed
parameters for Jupiter \citep[e.g,][]{hubb1977}; Saturn is hotter than
expected \citep{poll1977, stev1977, fort2003} and Uranus and Neptune
are colder than expected \citep{stev1982, podo1991, guil2005}. An
additional motivation to adopt a simple framework for temperature is
the huge uncertainty in interior parameters for rocky planets.  This
includes the temperature-dependent equation of state of liquids and
solids at high temperature and pressure, the temperature-dependent
viscosity, and the effect of tides on the cooling history for
eccentric exoplanets.  We further note that our choice of radiative
equilibrium down to 1 kbar is based on the irradiated
atmosphere/interior models by \citet{fort2007}.  For our fiducial
model we choose $T_{eff}$ to be similar to Earth's and Uranus's.
While the largest uncertainty in our treatment is the qualitative
choice of $T_{eff}$, we subsequently vary over a reasonable range of
effective temperatures (or adiabats).

We use the H/He equation of state from \citet{saum1995}, ignoring the
``plasma-phase transition'' which may be a numerical artifact (D. Saumon
2007, private communication). The equations of state for the solid
materials Fe \citep{ande2001}, MgSiO$_3$ perovskite \citep{kark2000},
(Mg,Fe)SiO$_3$ \citep{knit1987}, and H$_2$O are described in more
detail in \citet{seag2007}.

\section{Results and Discussion}

Our fiducial planet consists of a 30\% Fe core and a 70\% MgSiO$_3$
mantle, roughly analogous to Earth. We used a H/He mixture with helium
mass fraction $Y=0.28$ (the He mass fraction of the solar nebula).  We
chose $T_{eq}=300$~K, based on the observation that a planet around an
M-dwarf at an orbital distance of 0.1 AU has a similar equilibrium
temperature to Earth (assuming similar albedos). We set
$T_{eff}=30$~K, similar to Earth and Uranus\footnote{Earth has 44
$\times 10^{12}$~ W \citep{poll1993} and Uranus has $340 \times
10^{12}$~W of energy flow \citep{pear1990}.}. For the atmospheric
parameters, we fixed $\mu_0 = \cos 60^{\circ}$ and $\gamma=0.1$ to
represent radiation absorbed deep in the atmosphere\footnote{In
comparison $\gamma=10$ would correspond to absorption high in the
atmosphere.}. We later investigate variations on $Y$, $T_{eff}$,
$T_{eq}$ and $\gamma$.  Figure~1 shows a plot of the mass-radius
relationship for fiducial planets of masses 5, 10, 15, and 20
$M_{\oplus}$.  For each planet mass, we added atmospheres ranging in
mass from 0.001-1 $M_{\oplus}$.

A robust finding for all models is that a small amount of gas creates
a large radius increase.  While this result is expected, the radius
increase is far more dramatic than anticipated.

For example, an H/He atmosphere of $\sim 0.001$ by mass---only ten
times greater than Venus' atmospheric mass fraction---is required for
a noticeable radius increase. As seen in Figure~1, adding a
hydrogen-helium atmosphere with just 0.1\% of the mass of a 10
$M_{\oplus}$ rocky planet results in a 5\% increase in the planetary
radius---within a measurement precision that has been obtained for currently
known transiting planets.

\begin{figure}
\plotone{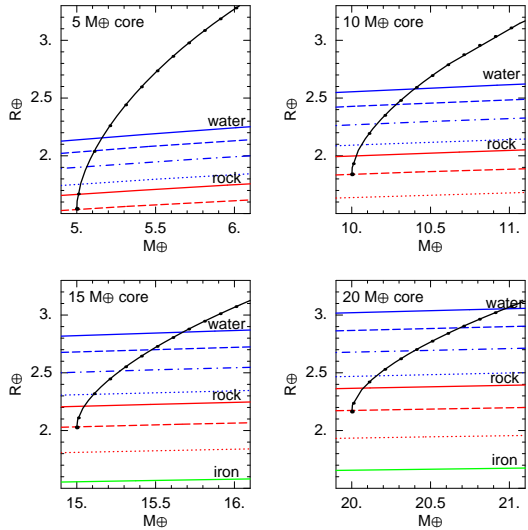}
\caption{The increase in radius due to adding H/He to a solid
planet. A H/He layer of 0.002-1 $M_{\oplus}$ is added to a solid
planet of 5, 10, 15, or 20 $M_{\oplus}$, with fiducial model
parameters (30\% Fe and 70\% MgSiO$_3$). The black points are for atmospheres
at 0.01 $M_{\oplus}$ and every 0.1 $M_{\oplus}$ afterwards. The
mass-radius relationship of solid planets with no gas is plotted for
comparison. The water (blue), rock (red), and iron (green) curves are
taken from \citet{seag2007} and represent homogeneous solid planets.
Intermediate compositions for differentiated planets are, from top
down: dashed-blue, 75\% H$_2$O, 22\% MgSiO$_3$, and 3\% Fe;
dashed-dotted-blue, 48\% H$_2$O, 48.5\% MgSiO$_3$, and 6.5\% Fe;
dotted-blue, 25\% H$_2$O, 52.5\% MgSiO$_3$, and 22.5\% Fe; dashed-red,
67.5\% MgSiO$_3$ and 32.5\% Fe; and dotted-red, 30\% MgSiO$_3$ and 70\%
Fe. In general, the addition of a gas layer of up to $\sim 5$\% of the solid
planet mass will inflate the radius of a rocky-iron planet through
the range of radii corresponding to water planets with different
water mass fractions. 
\label{fig:gas1}}
\end{figure}

As a second example, adding a gas layer of H/He equal to 1\% of the
mass of our fiducial planets increases the radius by $\sim$ 20\% of
the original planet radius, or by about 0.35 $R_{\oplus}$ for the
planet masses we considered. 

Our major finding is that that exoplanets with a significant H/He
layer cannot be distinguished from water worlds, based on $M_p$ and
$R_p$ alone.  For our fiducial solid exoplanets, adding up to 5\% H/He
by mass (for 10 $M_{\oplus}$ planets) is sufficient to push the
planet's radius through the entire range of radii corresponding to
solid planets with no gas, including planets with up to 100 percent
water composition. While we have not completed an exhaustive study of
possible compositions, we find the non-uniqueness of water planets to
be valid for any conditions we investigated.

This generic finding holds for a wide range of assumptions of assumed
temperatures.  Taking our fiducial model, we vary $T_{eq}$, $T_{eff}$,
and $\gamma$ individually.  For a 10 $M_{\oplus}$ solid planet with an
additional 0.1 $M_{\oplus}$ H/He atmosphere, increasing $T_{eq}$ from
300 to 500 K increases the radius by about ~1\% (Figure~2). For the
same planet varying $T_{eff}$ from 10~K to 50~K results in an 8\%
increase in radius. While large, this value is comparable
to expected radius uncertanties for these planets \citep{gill2007a,
gill2007b, demi2007}. Varying the altitude where radiation is absorbed
(specified by $\gamma$) has a much smaller effect on the planet
radius. Varying $\gamma$ from 0.1 to 10 causes the radius to decrease
by 0.2\%.

\begin{figure}
\plotone{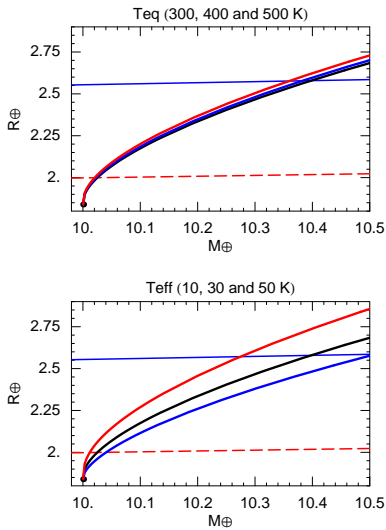}
\caption{The effects on the radius of varying the equilibrium
temperature (top) and effective temperature (bottom) for a 10 $M_{\oplus}$
planet with otherwise fiducial parameters. $T_{eq}$ values of 300, 400,
and 500~K are plotted to simulate the effect of uncertainty in the
orbital parameters and albedo on the expected radius. $T_{eff}$ values
of 10, 30, and 50 K are plotted on the same scale as the $T_{eq}$
plot, to show uncertainties in the planet's interior
temperature. Uncertainties in the internal energy of a planet lead to
large variations in radii for a given mass, showing how temperature is
a large uncertainty in the interpretation of a planet's internal
composition.
\label{fig:temp}}
\end{figure}

A corollary of our main result is that when a planet has a
significant H/He atmosphere there is a wide degeneracy in allowable
internal composition. This is not just compositional, but also relates
to the trade off of temperature and mass of H/He gas.  It could be
argued that specifying a planet's composition implies a particular
internal thermal profile derived from a consistent cooling history.
As addressed in \S2, the many unknowns and free input parameters for
rocky planet interiors---such as the possible differences between
atmospheric and interior compositions, equation of states, and the effect of
tides on the planet's cooling history---prevent a self-consistent
solution for the present time.

How could a 5--20 $M_{\oplus}$ exoplanet get a substantial H/He layer?
Two different scenarios may produce them: direct capture of gas from
the protoplanetary disk (possibly modified by the escape of some
fraction of the original gas) or outgassing during accretion.

A planet may capture and retain up to 1 to 2 $M_{\oplus}$ of H/He if
the planetary core did not grow quickly enough to capture more before
the gas in the disk evaporated (as is the paradigm for Uranus and
Neptune). Alternatively, for short-period exoplanets, a 1 to
2~$M_{\oplus}$ H/He envelope may result after substantial loss of an
initially massive gas envelope from irradiated evaporation
\citep[e.g.][]{bara2006}.  \citet{alib2006} consider atmospheric
evaporation during migration, and conclude that the $10 M_{\oplus}$
innermost planet in HD~69830, at 0.08 AU, kept $\sim$2 $M_{\oplus}$ of
H/He over the 4 Gyr lifetime of the star.

Little attention has been given to the mass and composition of
exoplanet atmospheres from outgassing. Venus' atmosphere is $10^{-4}
M_{\oplus}$; if Venus had a surface gravity high enough to prevent H
escape, its atmosphere would be over $10^{-3} M_{\oplus}$.  Even
more massive H-rich atmospheres are possible. If a
massive iron-silicate planet formed with enough water, the iron may
react with the water during differentiation, liberating hydrogen gas
\citep{ring1979, waen1994}.  L. Elkins-Tanton et al. (in prep.) estimate
that the maximum H component is about six percent by mass for a
terrestrial-composition planet. For a $10 M_{\oplus}$ planet this
would result in a $0.6 M_{\oplus}$ H envelope.

For short-period, low-mass planets, theoretical arguments of
atmospheric escape may be the best way to identify a water world based
on the mass and radius measurements alone \citep{lege2004,
sels2007}. Indeed, our assumption of H/He atmospheres for exoplanets
relies on the condition that atmospheric mass loss has not evaporated
all of the H/He. In the absence of hydrodynamic escape, the exospheric
temperature (and not the atmospheric $T_{eff}$) drives the thermal
Jeans escape of light gases. Earth and Jupiter both have exobase
temperatures of 1000~K \citep{depa2001}, significantly above their
$T_{eff}$ of 255~K and 124.4~K respectively \citep{cox2000}. Uranus
and Neptune have exobase temperatures around 750~K \citep{depa2001}.
We note that because GJ 436 must have at least 1~$M_{\oplus}$ of H/He,
its exospheric temperature is not too high.  On the other extreme,
planets of $5 M_{\oplus}$ would require very low exospheric
temperatures ($\sim 300$~K) to retain a massive atmosphere over the
course of billions of years. Nevertheless, a young $5 M_{\oplus}$
Earth-mass planet with a captured atmosphere could still have a H/He
atmosphere and an old $5 M_{\oplus}$ planet could retain a substantial
He fraction, making its compositional identification ambiguous.

\begin{figure}
\plotone{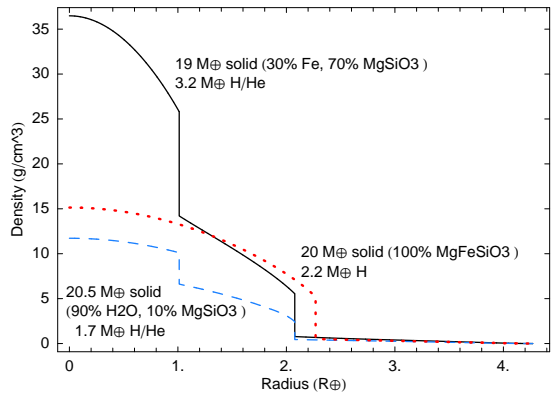}
\caption{Density vs. radius for three different potential compositions
of GJ436b. 
From top to bottom: solid (black) curve, 19.0 $M_{\oplus}$
core (30\% Fe, 70\% MgSiO$_3$) with 3.2 $M_{\oplus}$ H/He (Y=0.28);
dotted (red) curve, 20.0 $M_{\oplus}$ core (100\% (Mg,Fe)SiO$_3$) with 2.2
$M_{\oplus}$ H (Y=0); dashed (blue) curve, 20.5 $M_{\oplus}$ core
(90\% H$_2$O, 10\% MgSiO$_3$) with 1.7 $M_{\oplus}$ H/He (Y=0.28). All three
planets have the same total radius (4.3 $R_{\oplus}$) and total mass
(22.2 $M_{\oplus}$).
\label{fig:GJinterior}}
\end{figure}

We now turn to a qualitative study of GJ436b, to show that the
interpretation that GJ436b is a water world akin to Uranus and Neptune
\citep{gill2007b} is not the only possibility.  We consider the 
GJ~436b values $M_p =
22.2 M_{\oplus}$ and $R_p = 4.3 R_{\oplus}$ from \citet{demi2007}.
The internal structure in Figure~3 shows how three planets
with very different internal compositions can have the same total mass
and radius.  We first explore a planet similar to our fiducial model:
a 22.2 $M_{\oplus}$ solid planet with Earth-like iron/rock mass ratio
(30/70), $T_{eff}=30$~K, and $T_{eq}=600$~K in rough agreement with
the orbital parameters (assuming an albedo of 0.1). By adding
$\sim$3.2$M_{\oplus}$ of hydrogen-helium to the 19.0 $M_{\oplus}$
solid planet, we are able to reproduce GJ~436b's radius. We note that
the mass of gas is 15\% of the solid mass, likely too much to have
originated from outgassing, and so capture must be at least partially
invoked to explain such a massive atmosphere.  The second composition
for GJ~436b we considered is for water worlds, one with a 50\% water
mantle (by mass) and 50\% silicate core, and another with 90\% water
mantle and a 10\% silicate core. These planets also need some H/He to
match the known radius, 12\% and 8\% by mass,
respectively.  The third model approximates planets with atmospheres
created from outgassing, considering an extreme scenario where all of
the available water has oxidized iron, leaving a 100\% (Mg,Fe)SiO$_3$
solid planet core. To match the observed radius, a 22.2 $M_{\oplus}$
planet requires $\sim 2.2 M_{\oplus}$ of H alone, a case that assumes
no initial trapping and subsequent outgassing of He. The pure-hydrogen
atmosphere is 10\% of the mass of the solid planet. This is above the
theoretical maximum of outgassing based on observed abundances of
metallic iron in chondritic meteorites from our solar system (see
Elkins-Tanton et al. in prep).  Although not an exhaustive study,
the range of interior compositions illustrates the variety
of possibilities, though all models require some H/He.

While our study is preliminary, we make the robust point that H-rich
thick atmospheres will confuse the interpretation of planets based on
a measured mass and radius.  This point is independent of the
uncertainties retained by our model including $T_{eq}$, $T_{eff}$, the
mass fraction of H/He, and the mixing ratio of H and He. We find that
the identification of water worlds based on the mass-radius
relationship alone is impossible unless a significant gas layer can be
ruled out by other means. Spectroscopy is the most likely means and
may become routine with transit transmission and emission
spectroscopy, and eventually with spectroscopy by direct imaging.

\acknowledgements{We thank Mark Marley, Jonathan Fortney, and Wade
Henning for useful discussions.}

%---------------------------------------------------------------------------
%\bibliography{planets}

%---------------------------------------------------------------------------

\clearpage

\end{document}